 \documentstyle[eqsecnum,aps,prd]{revtex}
 \newcommand{\be}{\begin{equation}}
 \newcommand{\ee}{\end{equation}}
 \newcommand{\ba}{\begin{eqnarray}}
 \newcommand{\ea}{\end{eqnarray}}
 
 \def\d{\partial}

 \def\sqr#1#2{{\vcenter{\hrule height.#2pt
                        \hbox{\vrule width.#2pt height#1pt\kern#1pt
                        \vrule width.#2pt}
                        \hrule height.#2pt}}}
 \def\real{{\vrule height 1.6ex width 0.05em depth 0ex
     \kern -0.06em {\rm R}}}

 \def\PRD#1#2#3{{\it Phys.\ Rev.} {\bf D#1}, #2 (#3)}

 \def\NPB#1#2#3{{\it Nucl.\ Phys.} {\bf B#1}, #2 (#3)}
 
 \def\PLB#1#2#3{{\it Phys.\ Lett.} {\bf B#1}, #2 (#3)}

 \def\IJMPA#1#2#3{{\it Int.\ J.\ Mod.\ Phys.} {\bf A#1}, #2 (#3)}
 
 \def\MPLA#1#2#3{{\it Mod.\ Phys.\ Lett.} {\bf A#1}, #2 (#3)}
 
 \def\CQG#1#2#3{{\it Class.\ Quantum Grav.} {\bf #1}, #2 (#3)}
 
 \def\ANP#1#2#3{{\it Ann.\ Physics (N.Y.)} {\bf #1}, #2 (#3)}

\begin{document}
\title{A Solvable Model of Two-Dimensional Dilaton-Gravity
Coupled to a Massless Scalar Field}
\author{Marco Cavagli\`a\thanks{Electronic address:
cavaglia@aei-potsdam.mpg.de}}
\address{Max-Planck-Institut f\"ur Gravitationsphysik\\
Albert-Einstein-Institut\\
Schlaatzweg 1, D-14473 Potsdam, Germany}
\date{\today}
\maketitle
\begin{abstract}
We present a solvable model of two-dimensional dilaton-gravity coupled
to a massless scalar field. We locally integrate the field equations
and briefly discuss the properties of the solutions. For a particular choice
of the coupling between the dilaton and the scalar field the model can be
interpreted as the two-dimensional effective theory of 2+1 cylindrical
gravity minimally coupled to a massless scalar field.
\vskip 10truept
PACS number(s): 04.60.Kz, 04.20.Jb
\end{abstract}

\section{Introduction}
The investigation of lower-dimensional gravity is receiving a lot of
attention because of its connection with string theory,
dimensionally reduced models (minisuperspaces and midisuperspaces)
and black hole physics \cite{general}. Lower dimensional
models may further provide some insight into the difficult challenge of
quantizing gravity theories in the (more physical) four dimensional case.
Hence, many 0+1, 1+1 and 2+1 integrable models have been analyzed and
solved in the literature, both from the classical and quantum points of
view \cite{general2}.

In this context many papers have been devoted to the discussion of
integrable and non-integrable two-dimensional models \cite{general2}. It is
well known that two-dimensional dilaton-gravity with an arbitrary
potential is (classically) completely integrable \cite{filippov}. A
remarkable feature of this model is that any solution possesses a 
Killing vector \cite{ks}, i.e.\ the metric tensor and the dilaton $\phi$ can
be cast in the form \cite{filippov}
\ba
&&ds^2=4\rho(u,v)dudv\,,\label{metric}\\
&&\rho(u,v)=h(\psi)\d_u\psi\,\d_v\psi\,,~~~~~\phi\equiv\phi(\psi)\,,
\label{mstatic}
\ea
where $\psi$ is a harmonic function, i.e.\ $\d_u\d_v\psi=0$. (With a
somewhat improper terminology we will call these solutions ``static'',
even though the Killing vector is not timelike and hypersurface
orthogonal.) This is the content of the generalized Birkhoff theorem
\cite{filippov,ks}. When a scalar field is added to the model, the
Birkhoff theorem is no longer valid and non-static solutions appear. It
is then worthwile to investigate how the presence of a scalar field is
related to issues like integrability and absence of Killing vectors. This
perhaps can help shed light on some unsolved problems in classical and
quantum gravity as, for instance, the origin of the black hole entropy. 

Having this in mind, the purpose of this brief report is the discussion
of the general solution of the model described by the action 
\be
S=\int
d^2x\sqrt{-g}\,\left[\phi R+\gamma V(\phi) (\nabla\chi)^2\right]\,,
\label{action}
\ee
where $R$ is the Ricci scalar, $\phi$ is the dilaton field, and $\chi$ is
the massless scalar field. $\gamma$ is a coupling constant and we have
set $16\pi G$=1. Despite of the non trivial coupling between the dilaton
and the scalar field the model is completely solvable for a large 
class of functions $V(\phi)$. Even though in this simple model one cannot
obtain any black hole solution, the discussion of the general solution of
the model based upon Eq.\ (\ref{action}) is important at least for two
reasons: first this is (to our knowledge) the only known example of a
solvable dilaton-gravity-matter model with a non-trivial coupling between
the dilaton and the scalar field; second, when $V(\phi)=\phi$ the model
can be interpreted as the two-dimensional effective theory of 2+1
$\real\times S^1$ cylindrical gravity minimally coupled to a massless
scalar field \cite{ash}. In this case the dilaton plays the role of the
scale factor of $S^1$. 
\section{Field Equations and General Solutions\label{general}}
From the action Eq.\ (\ref{action}) it is straightforward to obtain the
field equations. They can be cast in a useful and simple form writing
the two-dimensional line element in the ``conformal gauge'' (\ref{metric}).
The result is
\ba
&&\d_u\d_v\phi=0\,,
\label{ephi}\\
&&\d_u\d_v(\ln\rho)=\gamma V'(\phi)\d_u\chi\d_v\chi\,,
\label{erho}\\
&&\d_u[V(\phi)\d_v\chi]+\d_v[V(\phi)\d_u\chi]=0\,,
\label{echi}\\
&&\rho\d_u\left({\d_u\phi\over\rho}\right)=\gamma V(\phi)(\d_u\chi)^2\,,
\label{constr1}\\
&&\rho\d_v\left({\d_v\phi\over\rho}\right)=\gamma V(\phi)(\d_v\chi)^2\,,
\label{constr2}
\ea
where the prime represents the derivative w.r.t.\ $\phi$. It is
surprising that the field equations can be locally integrated. The
equation for the field $\phi$ is the key for solving the system. From
Eq.\ (\ref{ephi}), it follows that $\phi$ is indeed a harmonic function.
Therefore, the general solution of Eq.\ (\ref{ephi}) is
$\phi(u,v)=a(u)+b(v)$, where $a(u)$ and $b(v)$ are arbitrary functions.
The general solution can be classified in three distinct classes: i)
$\phi(u,v)=\phi_0$, i.e.\ $\phi$ constant; ii) $\d_u\phi(u,v)=0$ or
$\d_v\phi(u,v)=0$, i.e.\ $\phi$ depending on a single variable; iii)
$\phi$ depending on both $u$ and $v$ variables. The first two cases
identify degenerate solutions of the model. Let us now discuss separately
the above cases. 
\subsection{Constant $\bf\phi$}
This is the simplest (degenerate) case. When $\d_u\phi(u,v)=0$ and
$\d_v\phi(u,v)=0$ Eqs.\ (\ref{erho}-\ref{constr2}) reduce to (we assume 
$V(\phi_0)\not=0$)
\be
\d_u\d_v(\ln\rho)=0\,,~~~~~
\d_u\chi=0\,,~~~~~\d_v\chi=0\,.\label{1-eq}
\ee
The general solution of Eqs.\ (\ref{1-eq}) is obvious
\be
\chi=\chi_0\,,~~~~~~~~~~\ln(\rho)=c(u)+d(v)\,,\label{1-sol}
\ee
where $\chi_0$ is a constant and $c(u)$ and $d(v)$ are two arbitrary
functions. Recalling Eq.\ (\ref{metric}) and using the reparametrization
invariance of the metric in the $u$ and $v$ variables, the
three-dimensional line element can be cast in the form $ds^2=4d\zeta
d\upsilon$, where $\zeta=\int du\, e^{c(u)}$ and $\upsilon=\int dv\,
e^{d(v)}$. The solution is then static and the spacetime is flat, as
expected since this case corresponds to a pure dilaton-gravity model with
vanishing potential. Since $\phi$ is constant, the above solution
represents a flat ($\real^2\times S^1$) spacetime with constant $S^1$
scale factor when interpreted as a 2+1 reduced model ($V(\phi)=\phi$). 
\subsection{$\bf \phi$ Depending on a Single Coordinate}
Let us suppose, without loss of generality, that $\d_v\phi=0$, i.e.\
$b(v)=0$. (Alternatively, $\d_u\phi=0$.) From Eq.\ (\ref{constr2}) we
have $\d_v\chi=0$, and Eqs.\ (\ref{erho}-\ref{constr1}) reduce to the
form 
\ba
&&\d_u\d_v(\ln\rho)=0\,,\label{2-rho}\\
&&{d^2\phi\over d u^2}-\d_u\ln(\rho) 
{d\phi\over du}=\gamma V(\phi)\left({d\chi\over
du}\right)^2\,.\label{2-chi}
\ea
(Equation (\ref{echi}) is identically satisfied.) The general solution of 
Eqs.\ (\ref{2-rho},\ref{2-chi}) is
\be
\rho={d\phi\over du}\exp\left[d(v)-\gamma\int^\phi_{\phi_0} 
d\phi'\,V(\phi')\left({d\chi\over d\phi'}\right)^2\right]\,,\label{2-sol}
\ee
where $d(v)$ is an arbitrary function of $v$, $\phi_0$ is an integration
constant, and $\chi\equiv\chi[\phi(u)]$. The two-dimensional line element
then represents a flat spacetime, even though the scalar field is not
constant. Note that all fields depend on a single variable ($\phi$),
however the solution (\ref{2-sol}) is not strictly a ``static'' solution in
the sense (\ref{metric},\ref{mstatic}). (A similar solution arises in
two-dimensional dilaton-gravity when the dilaton has vanishing
potential.) The nature of the solution becomes particularly
evident when $V(\phi')=\phi'$ and the model is interpreted in 2+1
dimensions. In this case the three-dimensional line element corresponding
to the solution (\ref{2-sol}) reads ($V(\phi')=\phi'$) 
\be
ds^2=4\exp\left[-\gamma\int^\phi_{\phi_0} d\phi'\,\phi'\left({d\chi\over
d\phi'}\right)^2\right]d\phi d\upsilon+\phi^2d\theta^2\,,\label{2-ds}
\ee
where we have defined the new coordinate $\upsilon$ as in the previous 
section and $\theta\in[0,1]$ is the $S^1$ variable. The geometry of
the spacetime is $\real^2\times S^1$, and the scale factor $\phi$ is a
function of the light cone variable orthogonal to $\upsilon$. For
instance, choosing $\chi=\chi_0+\arctan(\phi/k)$, Eq.\ (\ref{2-ds}) reads 
($x=\upsilon+\phi$ and $t=\upsilon-\phi$)
\be
ds^2=f_0\, e^{\gamma/2\over 
1+(x-t)^2/4k^2}(-dt^2+dx^2)+{(x-t)^2\over 4}d\theta^2\,.
\label{2-part}
\ee
\subsection{Complete Case}
In this case, setting
\be
\rho(u,v)=f(u,v)\,{da(u)\over du}\,{db(v)\over dv}\,,
\label{def-f}
\ee
and using the new coordinates $(a,b)$ (note that the coordinate
transformation is never degenerate, the degenerate cases being included
in the previous sections), equations (\ref{erho}-\ref{constr2})
reduce to 
\ba
&&\d_a\d_b(\ln f)=\gamma V'(a+b)\d_a\chi\d_b\chi\,,
\label{elnf}\\
&&V'(a+b)(\d_a\chi+\d_b\chi)+2V(a+b)\d_a\d_b\chi=0\,,
\label{echi2}\\
&&\d_a(\ln f)=-\gamma V(a+b)(\d_a\chi)^2\,,\label{constrf1}\\
&&\d_b(\ln f)=-\gamma V(a+b)(\d_b\chi)^2\,.\label{constrf2}
\ea
Now the system of second order partial differential equations
(\ref{elnf}-\ref{constrf2}) can be integrated, solving first Eq.\
(\ref{echi2}), and then using the solution $\chi$ in Eqs.\
(\ref{constrf1},\ref{constrf2}).  This program can be easily completed
using the new variables $z=a+b$ and $w=a-b$. Since Eq.\ (\ref{echi2}) and
the constraints (\ref{constrf1},\ref{constrf2}) imply Eq.\ (\ref{elnf}),
we can forget the latter and write 
\ba
&&\d^2_z\chi+{V'(z)\over V(z)}\d_z\chi=\d^2_w\chi\,,\label{zwchi}\\
&&\d_z(\ln f)=-\gamma V(z)
\left[(\d_z\chi)^2+(\d_w\chi)^2\right]\,,\label{zlnf}\\
&&\d_w(\ln f)=-2\gamma V(z)\d_z\chi\d_w\chi\,,\label{wlnf}
\ea
where $V'={dV(z)\over dz}$. Equation (\ref{zwchi}) corresponds to Eq.\
(\ref{echi2}) and Eqs.\ (\ref{zlnf}-\ref{wlnf}) are the sum and the
difference of Eqs.\ (\ref{constrf1},\ref{constrf2}) respectively.
Finally, given a solution of Eq.\ (\ref{zwchi}), the (logarithm of the)
physical conformal factor of the two-dimensional metric $f$ can be 
written as a functional of $\chi$ and locally cast in the form 
\ba
&&\ln f(z,w;f_0)=\ln f_0+\nonumber\\
&&~~~-2\gamma V(z)\int^w_{w_0}dw'\d_{w'}\chi(w',z)
\d_z\chi(w',z)+\label{lnf}\\
&&~~~-\gamma\int^z dz' V(z')\left\{\left[\d_w 
\chi(w,z')\right]^2+\left[\d_z
\chi(w,z')\right]^2\right\}_{w=w_0},\nonumber
\ea
where $f_0$ and $w_0$ are two constants. Let us now focus attention on
Eq.\ (\ref{zwchi}). Since the latter is a linear separable partial
differential equation in the variables $z$ and $w$, the solution
can be written in the form \cite{zwillinger}
\be
\chi(z,w)=\chi_0+\int_{-\infty}^\infty d\lambda\, 
C(\lambda)\,\eta(z,\lambda)\xi(w,\lambda)\,,
\label{gen-chi}
\ee
where $\eta$ and $\xi$ are the solutions of the linear second order 
ordinary differential equations
\be
{d^2\eta\over dz^2}+{V'(z)\over V(z)}{d\eta\over 
dz}=\lambda\eta\,,~~~~~{d^2\xi\over dw^2}=\lambda\xi\,.\label{ode}
\ee
It can be proved that the solution (\ref{gen-chi}) is the most general
solution of Eq.\ (\ref{zwchi}) provided that the completeness theorem for
both Eqs.\ (\ref{ode}) holds. (See for instance \cite{zwillinger}.) This
rather weak assumption is satisfied for a wide class of
physical, well-behaved, functions $V(z)$ in Eqs.\ (\ref{ode}). As a
concrete example of the formalism, let us now consider the case $V(z)=z$
corresponding to the 2+1-dimensional model. The (real) solution
(\ref{gen-chi}) reads (in this case it is straightforward to verify that
the completeness theorem holds because the first equation in (\ref{ode})
coincides with the Bessel/modified Bessel equation depending on the sign
of $\lambda$)
\be
\chi(z,w)=\chi_0+\int_0^\infty 
d\alpha\left\{\chi^{(1)}_\alpha(z,w)+\chi^{(2)}_\alpha(z,w)\right\}\,,
\label{real-gen-chi}
\ee
where
\ba
&&\chi^{(1)}_\alpha(z,w)=\bigl[A_1(\alpha)\sin(\alpha 
w)+B_1(\alpha)\cos(\alpha 
w)\bigr]\nonumber\\
&&~~~~~~~~~~~~~~~~\cdot\bigl[C_1(\alpha)J_0(\alpha 
z)+D_1(\alpha)Y_0(\alpha z)\bigr]\,,\label{chialpha1}\\
&&\chi^{(2)}_\alpha(z,w)=\bigl[A_2(\alpha)e^{\alpha
w}+B_2(\alpha)e^{-\alpha w}\bigr]\nonumber\\
&&~~~~~~~~~~~~~~~~\cdot\bigl[C_2(\alpha)I_0(\alpha
z)+D_2(\alpha)K_0(\alpha z)\bigr]\,,\label{chialpha2}
\ea
where $J_0$ and $Y_0$ are the zero order Bessel functions of first and
second kind, $I_0$ and $K_0$ are the zero order modified Bessel
functions, and the coefficients $A,..,D$ are real functions. 

Starting from Eqs.\ (\ref{real-gen-chi}-\ref{chialpha2}), or directly
from Eqs.\ (\ref{zwchi}-\ref{wlnf}), some interesting particular
solutions can be calculated. Let us look, for example, for a solution of
Eq.\ (\ref{zwchi}) of the form $\chi(z,w)=\zeta(z)+\upsilon(w)$.
Inserting the previous ansatz in Eq.\ (\ref{zwchi}), we have
\be
\chi(z,w)=\chi_0+k_1\ln z+k_2 w+k_3\left(w^2+{z^2\over 2}\right)\,.
\label{chi-pseudo}
\ee
The conformal scale factor of the two-dimensional line element is
\be
f=f_0\, z^{-\gamma k_1^2}g(z,w)\,,
\label{f-pseudo}
\ee
where
\be
g(z,w)=e^{-2\gamma w(k_1+k_3 z^2)(k_2+k_3 w)-{\gamma z^2\over 4}
(2k_2^2+4k_1 k_3+k_3^2 z^2)}\,.
\label{g-pseudo}
\ee
The solution (\ref{chi-pseudo}-\ref{g-pseudo}) is generally non-static in
the sense (\ref{metric},\ref{mstatic}) because the fields depend on both
variables. (A particular case is given by the choice $k_1=k_3=0$. In this
case $\chi\equiv\chi(w)$ and $f=\exp(-\gamma k_2^2 z^2/2)$ depends only
on $z$. Hence, according to (\ref{metric},\ref{mstatic}) we have a flat
spacetime but a non-static solution.) However, when $k_2=k_3=0$, i.e.\
$g(z,w)=1$, Eqs.\ (\ref{chi-pseudo}-\ref{g-pseudo}) reduce to the
(well-known) static solution \cite{mann}. (Alternatively, the latter can
be obtained directly from Eqs.\ (\ref{real-gen-chi}-\ref{chialpha2})
setting, for instance, $A_i=B_i=\delta(\alpha)$.) Let us see this in
detail. Recalling Eq.\ (\ref{metric}) and Eq.\ (\ref{def-f}) the
three-dimensional line element reads ($w=\tau$, $z=R$) 
\ba 
&&ds^2=f_0\, R^{-\gamma
k_1^2}(-d\tau^2+dR^2)+R^2d\theta^2\,,\label{ds-static}\\
&&\chi=k_1\ln(R/R_0)\,.\label{chi-static2}
\ea
The above solution can be cast in a more familiar form with a
redefinition of the integration constants and a change of coordinates.
Let us set $R=\beta r^{N/2}$ and $\tau=t M$, where $\beta$ is a parameter
with dimensions of length and $N$, $M$ are related to $k_1$ and $f_0$ by
$N=2/(1+4k_1^2)$, $M=(2/Nf_0)\beta^{1-2/N}$. With these redefinitions the
line element (\ref{ds-static}) becomes ($\gamma=-4$, low-energy string
case) 
\be
ds^2=-U(r)dt^2+\beta^2 U^{-1}(r) dr^2+\beta^2 
r^N d\theta^2\,,\label{ds-familiar}
\ee
where $U(r)=(2M/N)r^{1-N/2}$. As expected, the model does not admit
black hole solutions and the metric becomes $\real^3$ flat when
$k_1=0$, i.e.\ when the scalar field is constant. This can be verified
directly from Eq.\ (\ref{lnf}), recalling that $\phi$ is a harmonic 
function, and Eq.\ (\ref{def-f}).

Let us briefly conclude the section with another interesting set of
solutions. Choosing for instance in Eq.\ (\ref{chialpha1}) $A_1=D_1=0$, $B_1
C_1=K$, and using Eq.\ (\ref{real-gen-chi}), we have \cite{bateman} 
\be
\chi(z,w)=\chi_0+{K\over\sqrt{z^2-w^2}}\,,\label{chi-dyn}
\ee
where $K$ is a constant. From Eq.\ (\ref{lnf}) it is straightforward to obtain 
the metric
\be
ds^2=f_0\, e^{\gamma K^2 z^2\over 2(z^2-w^2)^2}(-dw^2+dz^2)+z^2 
d\theta^2\,.
\label{metr-dyn}
\ee
The choice $B_1=D_1=0$, $A_1C_1=K$ gives instead the complementary solution
\ba
&&\chi(z,w)=\chi_0+{K\over\sqrt{w^2-z^2}}\,,\label{chi-dyn2}\\
&&ds^2=f_0\, e^{-{\gamma K^2 z^2\over 2(w^2-z^2)^2}}(-dw^2+dz^2)+z^2
d\theta^2\,.
\label{metr-dyn2}
\ea
The discussion of the global properties of Eqs.\
(\ref{chi-dyn}-\ref{metr-dyn2}) is beyond the scope of this brief report
and will be discussed elsewhere, so here we will not enter into details.
Let us stress, however, that the above solutions are asymptotically flat
and singular when $w\pm z=0$, i.e.\ when the scalar field diverges. With
different choices of the coefficients in Eqs.\
(\ref{chialpha1}-\ref{chialpha2}) is it possible to construct non
singular solutions. 
\section{Conclusions\label{concl}}
In this note we have briefly discussed the two-dimensional
dilaton-gravity-matter theory described by the action (\ref{action}).
When $V(\phi)=\phi$ the model can be interpreted as the two-dimensional
effective theory of 2+1 cylindrical gravity minimally coupled to a
massless scalar field. 

The model of Eq.\ (\ref{action}) has the remarkable property of being
completely solvable and we have derived and classified its solutions for
a large class of functions $V(\phi)$. It is well-known that any
two-dimensional pure dilaton-gravity theory satisfies the generalized
Birkhoff theorem, i.e.\ any solution of the system can be reduced to the
form (\ref{metric}-\ref{mstatic}). Here, due to the presence of the
scalar field, the Birkhoff theorem is no longer valid. It is then
interesting to examine how the presence of the scalar field modifies the
equations of motion w.r.t. to the pure dilaton-gravity case. Let us
consider for simplicity $\d_u\phi\not=0$ and $\d_v\phi\not=0$. In this
case Eqs.\ (\ref{ephi}) and (\ref{echi}-\ref{constr2}) imply Eq.\
(\ref{erho}), so we can neglect the latter. Equation (\ref{echi}) simply
defines the scalar field and can be solved setting
$V(\phi)\d_v\chi=\d_v\varphi$ and $V(\phi)\d_u\chi=-\d_u\varphi$. Using
$\varphi$ the remaining equations read 
\ba
&&\d_u\d_v\phi+\rho \bar V(\phi)=0\,,
\label{ephic}\\
&&\rho\d_u\left({\d_u\phi\over\rho}\right)=\gamma 
V^{-1}(\phi)(\d_u\varphi)^2\,, \label{constr1c}\\
&&\rho\d_v\left({\d_v\phi\over\rho}\right)=\gamma 
V^{-1}(\phi)(\d_v\varphi)^2\,, \label{constr2c}
\ea
where we allow for the presence of a dilatonic potential $\bar V(\phi)$.
Equation (\ref{ephic}) does not depend on $\varphi$. Hence, only the
constraints (\ref{constr1c},\ref{constr2c}) are modified by the presence
of the scalar field. The validity of the Birkhoff theorem is thus
related to the r.h.s.\ of the Eqs.\ (\ref{constr1c},\ref{constr2c}).
Further, a static solution is obtained only when the r.h.s.\ of Eq.\ 
(\ref{constr1c}) is equal to the r.h.s.\ of Eq.\ (\ref{constr2c}). In that
case $\varphi\equiv\varphi(u+v)$ or $\varphi\equiv\varphi(u-v)$ and the
equations of motion can be reduced to a system of ordinary differential
equations. Clearly, the above condition is satisfied in the case of pure
dilaton-gravity when $\varphi$ is identically zero but cannot be
satisfied by any dilaton-gravity theory coupled to a scalar field. 
Finally, the integrability property of the system does not seem to be
related to the existence of non-static solutions. 
\acknowledgements
We are indebted to Vittorio de Alfaro, Alexandre T.\ Filippov, Larry
Ford, and Alexander Vilenkin for interesting discussions and useful
suggestions on various questions connected to the subject of this paper.
This work has been partially supported by the grant No.\ 3229/96 of the
University of Torino and by a Human Capital and Mobility grant of the
European Union, contract number ERBFMRX-CT96-0012. 

\end{document}